\documentclass[journal=jacsat,manuscript=article,layout=twocolumn]{achemso}

\setkeys{acs}{
abbreviations=true,
articletitle=true,
chaptertitle=true,
doi=true,
email=true,
etalmode=truncate,
keywords=true,
maxauthors=100,
super=true
}

\usepackage[version=4]{mhchem}
\usepackage{booktabs}
\usepackage{threeparttable}
\usepackage[binary-units = true]{siunitx}
\usepackage{xcolor}
\usepackage{cuted}
\usepackage{{ragged2e}}
\usepackage[none]{hyphenat}
\usepackage{microtype}
\usepackage{array}
\usepackage{multicol} 
\usepackage{rotating}
\usepackage{subfigure}

\DeclareCaptionLabelFormat{plain}{#2}

\usepackage[colorlinks=true,allcolors=black]{hyperref}
\hypersetup{colorlinks=true,citecolor=blue,urlcolor=blue}
\newcommand{\doi}[1]{\href{http://dx.doi.org/#1}{\nolinkurl{#1}}}

\SectionsOn
\SectionNumbersOn
\AbstractOn
\vfuzz \hfuzz
\author{Rafael Maglia de Souza}
\affiliation{Departamento de Qu{\'i}mica Fundamental, Instituto de Qu{\'i}mica, Universidade de S{\~a}o Paulo, Avenida Professor Lineu Prestes \textit{748}, S{\~a}o Paulo, SP, Brasil}
\email{rafael.maglia.souza@usp.br}

\author{Mikko Karttunen}
\affiliation{Department of Chemistry, The University of Western Ontario, $1151$ Richmond Street, London, Ontario, Canada, N6A{~}5B7}
\alsoaffiliation{Department of Physics and Astronomy, The University of Western Ontario, $1151$ Richmond Street, London, Ontario, Canada N6A{~}3K7}
\alsoaffiliation{Centre for Advanced Materials and Biomaterials Research, The University of Western Ontario, $1151$ Richmond Street, London, Ontario, Canada, N6A{~}5B7}
\alsoaffiliation{Institute of Macromolecular Compounds, Russian Academy of Sciences, Bolshoi
Prospect V.O. 31, St.Petersburg 199004, Russia}
\email{mkarttu@uwo.ca}

\author{Mauro Carlos Costa Ribeiro}
\affiliation{Departamento de Qu{\'i}mica Fundamental, Instituto de Qu{\'i}mica, Universidade de S{\~a}o Paulo, Avenida Professor Lineu Prestes \textit{748}, S{\~a}o Paulo, SP, Brasil}
\email{mccribei@iq.usp.br}

\title{Fine-Tuning the Polarizable CL\&Pol Force Field for the Deep Eutectic Solvent Ethaline}


\begin{document}

\begin{tocentry}
\includegraphics[width=0.80\linewidth]{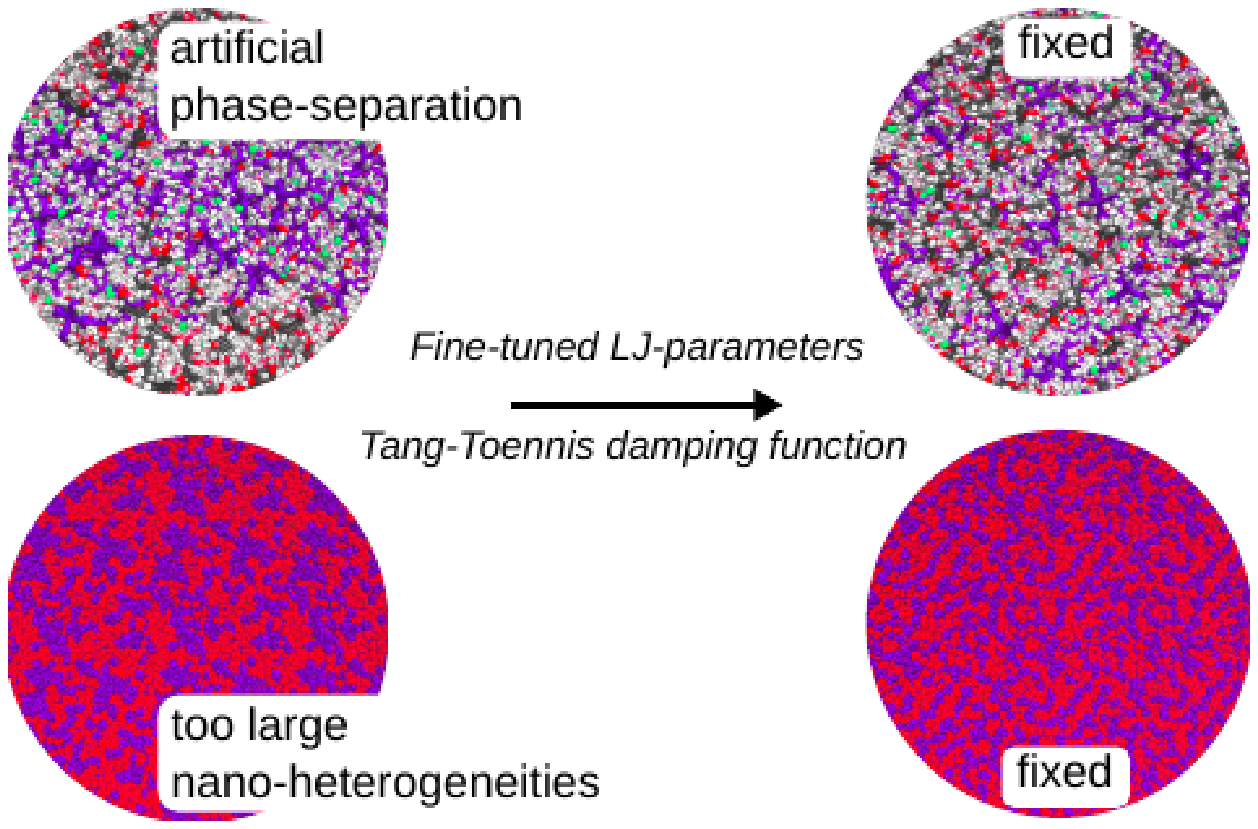}
\end{tocentry}

\begin{abstract}

Polarizable force fields are gradually becoming a common choice for ionic soft matter, in particular for molecular dynamics (MD) simulations of ionic liquids (ILs) and  deep eutectic solvents (DESs). The CL\&Pol force field introduced in 2019 is the first general, transferable and polarizable force field for MD simulations of different types of DESs. The original formulation contains, however, some problems that appear in simulations of ethaline and may also have a broader impact. First, the originally proposed atomic diameter parameters are unbalanced, resulting in too weak interactions between the chlorides and the hydroxyl groups of the ethylene glycol molecules. This, in turn, causes an artificial phase separation in long simulations. Second, there is an overpolarization of chlorides due to strong induced dipoles that give rise to the presence of peaks and antipeaks at very low $q$-vector values (\SI{2.4}{\per\nano\meter}) in the partial components of the structure factors. In physical terms, this is equivalent to  overestimated spatial nano-scale heterogeneity. To correct these problems, we adjusted the chloride-hydroxyl radial distribution functions against \textit{ab initio} data and then extended the use of the Tang-Toennis damping function for the chlorides' induced dipoles. These adjustments correct the problems without losing the robustness of the CL\&Pol  force field. The results were also compared with the non-polarizable version, the CL\&P force field. We expect that the corrections will facilitate reliable use of the CL\&Pol  force field for other types of DESs. 

\end{abstract}

\section{Introduction}

Deep eutectic solvents\cite{Smith2014, Hansen2020} (DESs) are an interesting class of materials resembling ionic liquids\cite{Dong2017, Singh2020} (ILs) with great potential applications in synthesis\cite{Carriazo2012}, electrochemistry\cite{Wu2021}, extraction processes\cite{Cunha2018}, and biomass transformation\cite{Chen2019}. Similarly, most DESs and ILs are composed of a molecular cation and an inorganic anion. However, DESs also have a neutral component in their compositions, such as glycol that forms a specific stoichiometric mixture\cite{Alizadeh2020, Martins2018}. ILs, on the other hand, are pure molten salts at (or near) room temperature. The components of DESs are typically classified as hydrogen bond donors (HBD) or hydrogen bond acceptors (HBA). The combined action\cite{Kaur2020} of all hydrogen bonds present in DESs contributes to a decrease in the eutectic point of the system in relation to what would be predicted for an ideal mixture\cite{Smith2014, Hansen2020, Martins2018}.  

In terms of MD simulations,
ILs and DESs present common characteristics and similar problems\cite{Kaur2020, GonzlezdeCastilla2019, Bedrov2019}. Due to the importance of electrostatic interactions in these systems, assigning partial charges is not straightforward. Several charge assignment methods exist and it is well-known that the choice impacts the properties and behavior of the systems\cite{Garca2015, Kohagen2011, Dommert2010, Schmidt2010, Schrder2008}. In addition, the high concentration of ions in these materials results in non-negligible local electric fields that polarize the voluminous ions. Due to this, polarization should be included for the simulations to be predictive and accurate. This is usually accomplished by either explicitly including polarization in force fields or by an implicit inclusion through a mean-field approach\cite{Bedrov2019, Cieplak2009,Lemkul2016}.

Polarization can be implicitly accounted for, e.g., by modifying the parameters of the Lennard-Jones (LJ) potential. This approach is based on the fact that the interaction between
two induced dipoles depends on the interparticle distance ($r$) as $\sim \! \! r^{-6}$, the same dependence that is present in the dispersion term of the LJ potential. This approach has been used for both DESs\cite{Chaumont2020} and ILs\cite{Kddermann2007}. The problems include the computational effort to recalibrate the LJ parameters, poor transferability, the absence of directionality and difficulties in accurately describing transport properties\cite{Bedrov2019}.

A more popular and easy way for implicit inclusion of polarization is re-scaling the fixed partial charges of the atoms, which has been extensively done for both ILs and DESs  across different modeling resolutions\cite{Zhang2020, Perkins2014, deSouza2021, Sapir2020, Celebi2019, deSouza2019, Salanne2015}. The theoretical framework for reducing charges is the electronic continuum correction (ECC) proposed by Leontyev and Stuchebrukhov\cite{Leontyev2009, Leontyev2011}. They showed that charges should be scaled by a multiplicative factor $\gamma = 1/n$, where $n$ is the refractive index of the medium; it is connected to the optical dielectric constant, $\epsilon_{\infty}$,  via $\epsilon_{\infty} = n^{2}$. The scaling factor is typically in the range of \num{0.7} - \num{0.9}. It provides an approach consistent with quantum-mechanical charge evaluations and serves as an adjustable parameter for calibrating force fields\cite{Sapir2020} but it does not, however, come without problems\cite{Tolmachev2020-rw}.

Explicit inclusion of polarization is currently a very active field of research\cite{Bedrov2019, Lemkul2016}. The methods include induced point dipoles, fluctuating charge models and the classical Drude oscillator. The main drawback of polarizable force fields is the large computational cost, making their development and calibration a challenging task. Despite the challenges, several general polarizable force fields exist for ILs, such as APPLE$\&$P\cite{Borodin2009}, CL\&Pol\cite{Goloviznina2019} and AMOEBA\cite{Starovoytov2014}. For DESs, the available option is the CL\&Pol force field by Goloviznina \textit{et al.}\cite{Goloviznina2019, Goloviznina2021} based on Drude-induced dipoles. 

\begin{figure}[b!]
\centering
\includegraphics[width=1.00\linewidth]{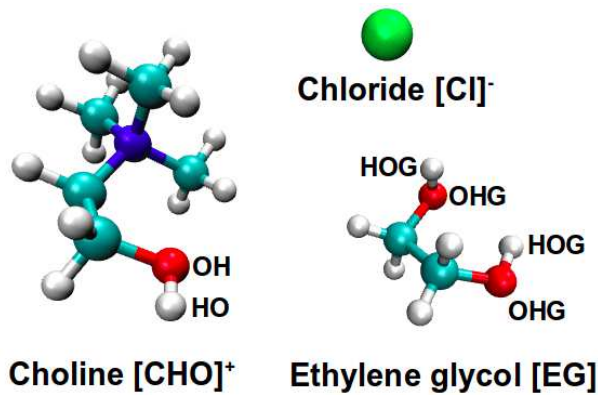}
\caption{Chemical structures of DES components and the naming conventions for the atom types of the hydroxyl groups. Oxygens are depicted in red, hydrogens in white, carbons in cyan, nitrogen in blue, and chloride in green.}
\label{fig:structures}
\end{figure}

So far, only general validation has been completed with the CL\&Pol force field for DES. This includes properties such as density, viscosity and diffusion coefficients\cite{Goloviznina2021}.  More complex evaluations of liquid structure and dynamic behavior of common DESs have been done with traditional non-polarizable force fields, such as the OPLS\cite{Doherty2018, Bittner2021, CeaKlapp2021}, CHARMM/CGenFF\cite{Kaur2019, JahanbakhshBonab2021, Bittner2021} and AMBER/GAFF\cite{Zhang2020, Mainberger2017, Bittner2021} families. Of these force fields, only OPLS has been specifically tuned for DES\cite{Doherty2018}. Other works that have addressed calibration or adjustment of parameters, includes those of Perkins\cite{Perkins2013, Perkins2014} \textit{et al.} and Ferreira\cite{Ferreira2016} \textit{et al.}

In the present study, we explored the application of the CL\&Pol force field\cite{Goloviznina2021} and we noted some important issues that lead to unphysical results: Phase separation and overpolarization that occurs with the chlorides. Here, we present, discuss, and correct both problems, which allows a safer application of the force field. In addition, we also compared the results with its reduced charge non-polarizable version, the CL$\&$P force field\cite{CanongiaLopes2012}. The chosen DES was composed of the choline chloride, \ce{[CHO]^+}\ce{[Cl]^-}, and ethylene glycol, \ce{[EG]}, at a molar ratio of \num{1}:\num{2}, called ethaline, which has been subject of several experimental and theoretical studies\cite{Alizadeh2020, Kaur2020, Kaur2019, Zhang2020, Wagle2016, Ferreira2016}. Figure{~}\ref{fig:structures} shows the chemical structures and nomenclature of atom types used in this work. 

\section{Computational Details}

\subsection{CL{\&}P and CL{\&}Pol Force Fields}\label{sec:dev}

The CL{\&}P force field\cite{CanongiaLopes2012} was originally developed using fixed charges in a non-polarizable fashion. It was further extended to include explicit polarization via Drude-induced dipoles and named as CL{\&}Pol\cite{Goloviznina2019, Goloviznina2021}. CL{\&}Pol shares the  functional form of the CL{\&}P force field, which is composed of Coulomb and Lennard-Jones (LJ) non-bonded interactions, and bond, angle and torsional potentials for intramolecular interactions. In addition, for modeling DESs, the CL{\&}Pol force field also has the Tang-Toennies\cite{Tang1984} (TT) and Thole\cite{Thole1981} damping functions, 
\begin{center}
\begingroup
  \small   
  \thinmuskip=\muexpr\thinmuskip*1/9\relax
  \medmuskip=\muexpr\medmuskip*1/9\relax 
\begin{equation}\label{eqn:TT}
\begin{split}
f_{n}(r_{ij}) = \num{1} - ce^{-br_{ij}} \sum_{k=0}^{n}\frac{(br_{ij})}{k!}^{k}, \\ \end{split}
\end{equation}
\endgroup\\
\end{center}
\begin{center}
\begingroup
  \small   
  \thinmuskip=\muexpr\thinmuskip*1/9\relax
  \medmuskip=\muexpr\medmuskip*1/9\relax 
\begin{equation}\label{eqn:Thole}
\begin{split}
T(r_{ij}) = \num{1} - \left(\num{1} + \frac{pr_{ij}}{\num{2}(\alpha_{i}\alpha_{j})^{\frac{1}{6}}} \right) e^{-pr_{ij}/{(\alpha_{i}\alpha_{j})^{\frac{1}{6}}}}, \\ \end{split}
\end{equation}
\endgroup\\
\end{center}
where the parameter $b$ determines the 
spatial extension of damping (the developers assigned its value equal to \num{4.5}), $k$ is the order of the sum and goes to $k$=\num{4}, $c$=\num{1}, $r_{ij}$ is the distance between the sites, $p$=\num{2.6} is the Thole Parameter and $\alpha_{i}$ is the polarizability of atom $i$. 

The Thole function dampens the Coulomb interactions at short distances that originate from the induced dipoles. The TT function was adopted in the CL{\&}Pol force field to avoid instabilities and to dampen the short-range charge-dipole interactions that occur due to the presence of "naked" hydrogen atoms in the force field formulation\cite{Goloviznina2021}. It means that some hydrogen atoms do not have LJ parameters, just charges.

The "naked" hydrogens give rise to an important issue: The atomic diameters (the $\sigma$-parameter of the LJ potential)
must be modified for the cross-interactions between chlorides and hydroxyls of the \ce{[CHO]^+} and \ce{[EG]} species. The motivation is that the repulsive potentials of the oxygen atoms (\ce{OH} and \ce{OHG}, Figure{~}\ref{fig:structures}) to which the hydrogens from hydroxyls are bound, are insufficient to compensate the augmented polarization effect, which "freezes" the system and impacts the local structure. They proposed\cite{Goloviznina2021} the value of $\sigma$=\SI{0.37}{\nano\meter} for both $\ce{[Cl]^-}-\ce{OH}$ and $\ce{[Cl]^-}-\ce{OHG}$ interactions. As we will see in Section~\ref{sec:finetune}, that leads to an unphysical phase separation. The fine-tuned values we propose are $\sigma$=\SI{0.345}{\nano\meter} and $\sigma$=\SI{0.356}{\nano\meter} for the $\ce{[Cl]^-}-\ce{OHG}$ and $\ce{[Cl]^-}-\ce{OH}$, respectively.

The CL{\&}Pol force field assigns all the force constants of the harmonic bonds between the Drude cores (DC) and the Drude particles (DP) to $k_{D}$=\SI{4184}{\kilo\joule\per\mol}, and masses of all DPs to $m_{DP}$=\num{0.4}\,u. In addition, the CL{\&}Pol force field treats only the heavy atoms as polarizable and adds the polarizabilities of hydrogen atoms to the heavy atom to which they are bonded.  Polarizabilities are used to generate the partial charges ($q_D$) of the DPs ($\alpha$=$q_D^{2}$/$k_{D}$) and have negative values. Opposite charges (-$q_D$) are added onto the initial charges of the heavy atoms.

These \textit{starting} charges come from the non-polarizable CL{\&}P force field as do the \textit{starting} LJ parameters. The LJ parameters are subsequently scaled to avoid double counting of polarization effects\cite{Goloviznina2021}. This process is based on a predictive scheme proposed by the developers of the original force field\cite{Goloviznina2019, Pdua2017} using key fragments of the molecules of interest to avoid costly symmetry-adapted perturbation theory (SAPT) calculations\cite{Szalewicz2012-pf}. They split the \ce{[EG]} into two methanol units, while the cholinium cation was treated as a single fragment. The scaling coefficients and the final set of LJ parameters (as well as the partial charges and intramolecular parameters) they obtained were used in our work and can be consulted in their publication\cite{Goloviznina2021}, except for the $\sigma$ parameter of the $\ce{[Cl]^-}-\ce{OHG}$ and $\ce{[Cl]^-}-\ce{OH}$ interactions, as mentioned above. 

In our work, we also performed MD simulations with the non-polarizable CL{\&}P force field. In this case, the partial charges of the \ce{[CHO]^+}\ce{[Cl]^-} was multiplied by a scaling factor equal to \num{0.9} to implicitly account for polarization. The simulation with this scaling factor was validated by the reproduction of physicochemical properties of ethaline, as will be seen in Section~\ref{sec:finetune} and Table~\ref{tbl:FF}.

\subsection{Molecular Dynamics Simulations}

All the MD simulations were executed using LAMMPS\cite{Plimpton1995} with periodic boundary conditions and with the USER-DRUDE package\cite{Dequidt2015} enabled to allow the use of Drude-induced dipoles. To integrate the equations of motion, time steps of \SI{1}{\femto\second} and \SI{2}{\femto\second} were employed in the polarizable and non-polarizable simulations, respectively. A cut-off radius of \SI{1.2}{\nano\metre} was used for the Lennard-Jones and the real space part of the electrostatic interactions. The particle-particle particle-mesh (P3M) method\cite{Hockney1973, Eastwood1975} was applied for the long-range part of the electrostatic interactions. Bonds connected to hydrogen atoms were constrained using the SHAKE algorithm\cite{Ryckaert1977}.

The ethaline simulations were conducted using \num{150} \ce{[CHO]^+}\ce{[Cl]^-} ionic pairs and \num{300} \ce{[EG]} molecules (\num{1}:\num{2} molar ratio). Initial configurations and force field parameter assignment were performed with \textit{fftool} and Packmol\cite{Martnez2009} utilities. In addition, for the polarizable model, the \textit{polarizer} and \textit{scaleLJ} tools were applied to add the Drude particles and to scale the Lennard-Jones parameters following the CL{\&}Pol force field protocol\cite{Goloviznina2021}. 

The steepest descents algorithm was used for energy minimization of the initial configurations. Then, the systems were equilibrated for \SI{10}{\nano\second} in the NPT (P=\SI{1}{atm} and T=\SI{323.15}{\kelvin} or \SI{373.15}{\kelvin}) ensemble and subsequently simulated in the NVT ensemble for \SI{70}{\nano\second}. In the non-polarizable simulation, temperature and pressure were controlled using the Nosé-Hoover thermostat and barostat\cite{Nos1983, Hoover1985} with time constants of \SI{0.2}{\pico\second} and \SI{2}{\pico\second}, respectively. In the simulations with polarizable force field, temperature and pressure were controlled using the temperature-grouped Nosé-Hoover thermostat\cite{Son2019} to provide better kinetic energy equipartitioning
\cite{Son2019}. The time constants were set to \SI{0.1}{\pico\second} and \SI{1}{\pico\second} for temperature and pressure, respectively. In addition, the temperature of the Drude particles was maintained at \SI{1}{\kelvin} with time constant of \SI{0.02}{\pico\second}.

To obtain the X-ray structure factors, $S(q)$, the last frame from each of the production runs was taken and replicated twice
in the $x$, $y$, and $z$-directions. The final box consisted of \num{1200} \ce{[CHO]^+}\ce{[Cl]^-} ionic pairs and \num{2400} \ce{[EG]} molecules. It was equilibrated for \SI{2}{\nano\second} in the NPT ensemble and simulated for \SI{5}{\nano\second} in the NVT ensemble. The larger simulation box was needed to achieve low $q$ values in the S(q) calculations.

\begin{figure*}[t!]
\centering
\includegraphics[width=1.00\linewidth]{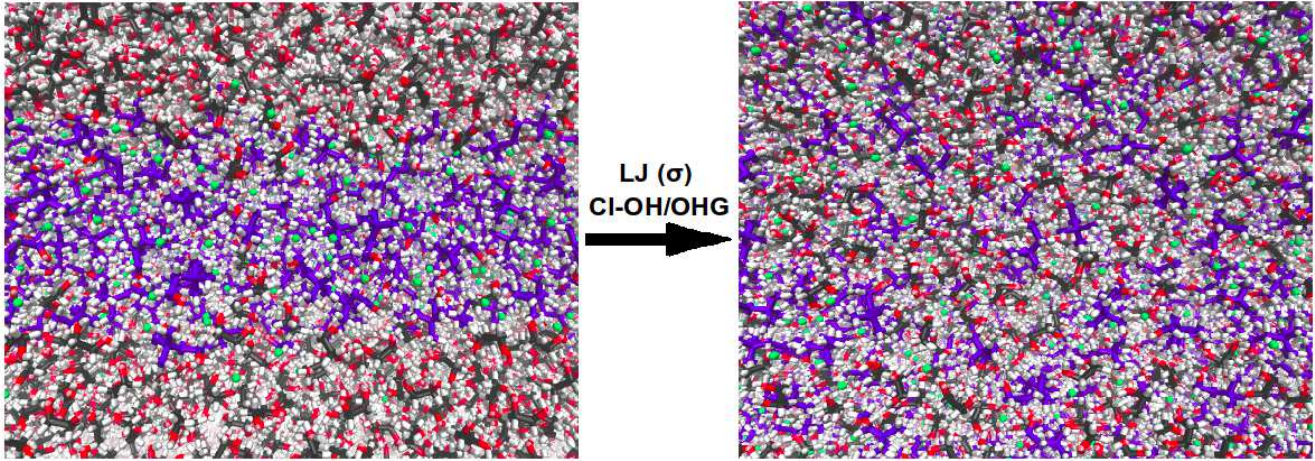}
\caption{Images showing the unphysical phase separation (left) and the behavior after fine-tuning (right) of the CL{\&}Pol force field, both at \SI{323.15}{\kelvin}. Choline ions are shown in blue, ethylene glycol molecules in black, and chlorides in green. Oxygens are highlighted in red and hydrogens in white.}
\label{fig:phase}
\end{figure*}

\section{Results and Discussion}

We start demonstrating and discussing the unphysical phase separation that occurs when using the original setup of the CL{\&}Pol force field. This is followed by a proposed correction to the problem. Then, we present evidence and discuss the overpolarization of chloride anions, followed by our proposal to correct it, namely the application of the TT damping function to the induced dipoles of the chlorides. In both sections, we also provide a comparison with the non-polarizable CL{\&}P force field.

\subsection{Artificial Phase Separation and How to Correct it
\label{sec:finetune}}

\begin{figure}[b!]
\centering
\includegraphics[width=1.00\linewidth]{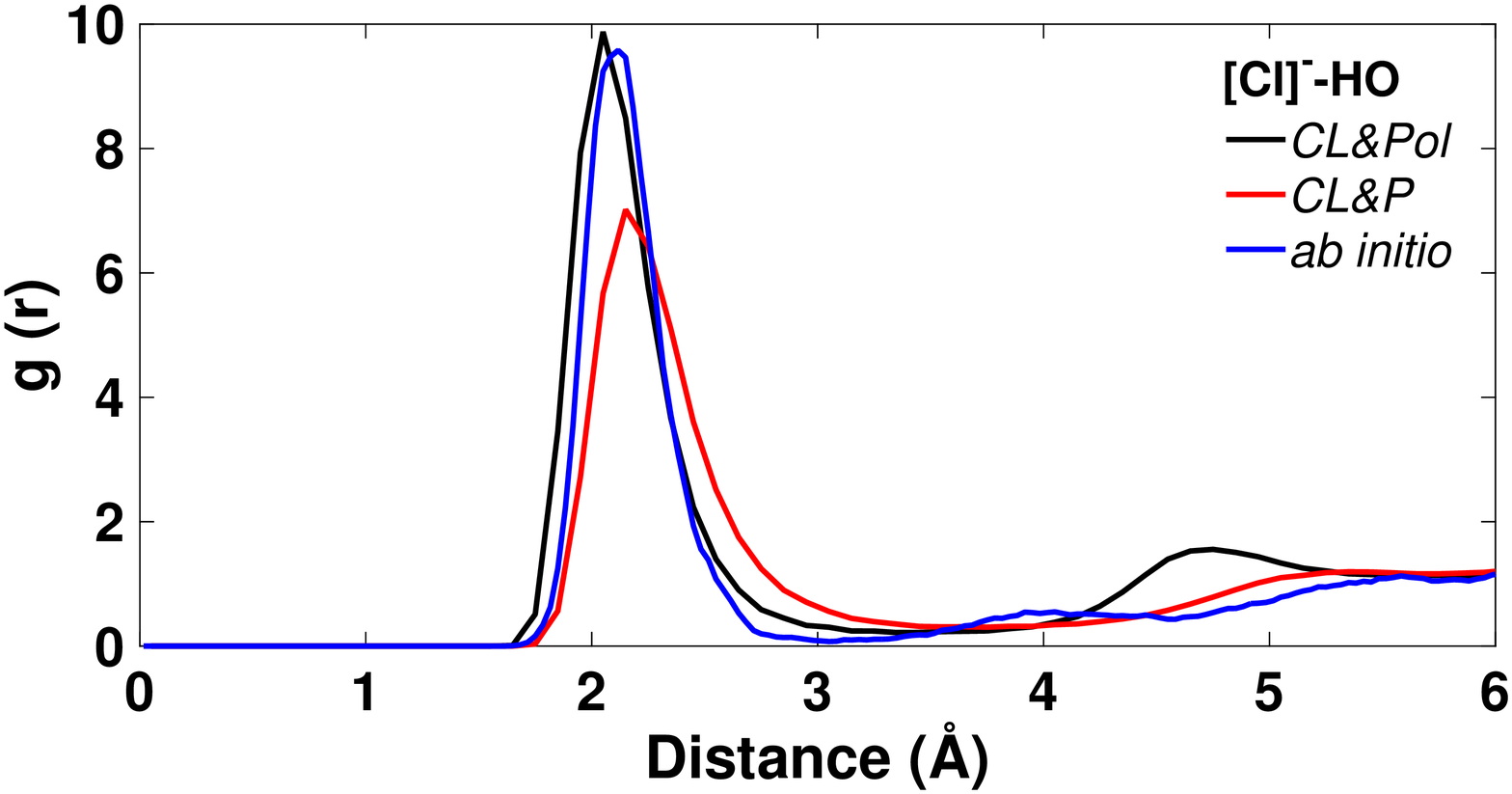}
\includegraphics[width=1.00\linewidth]{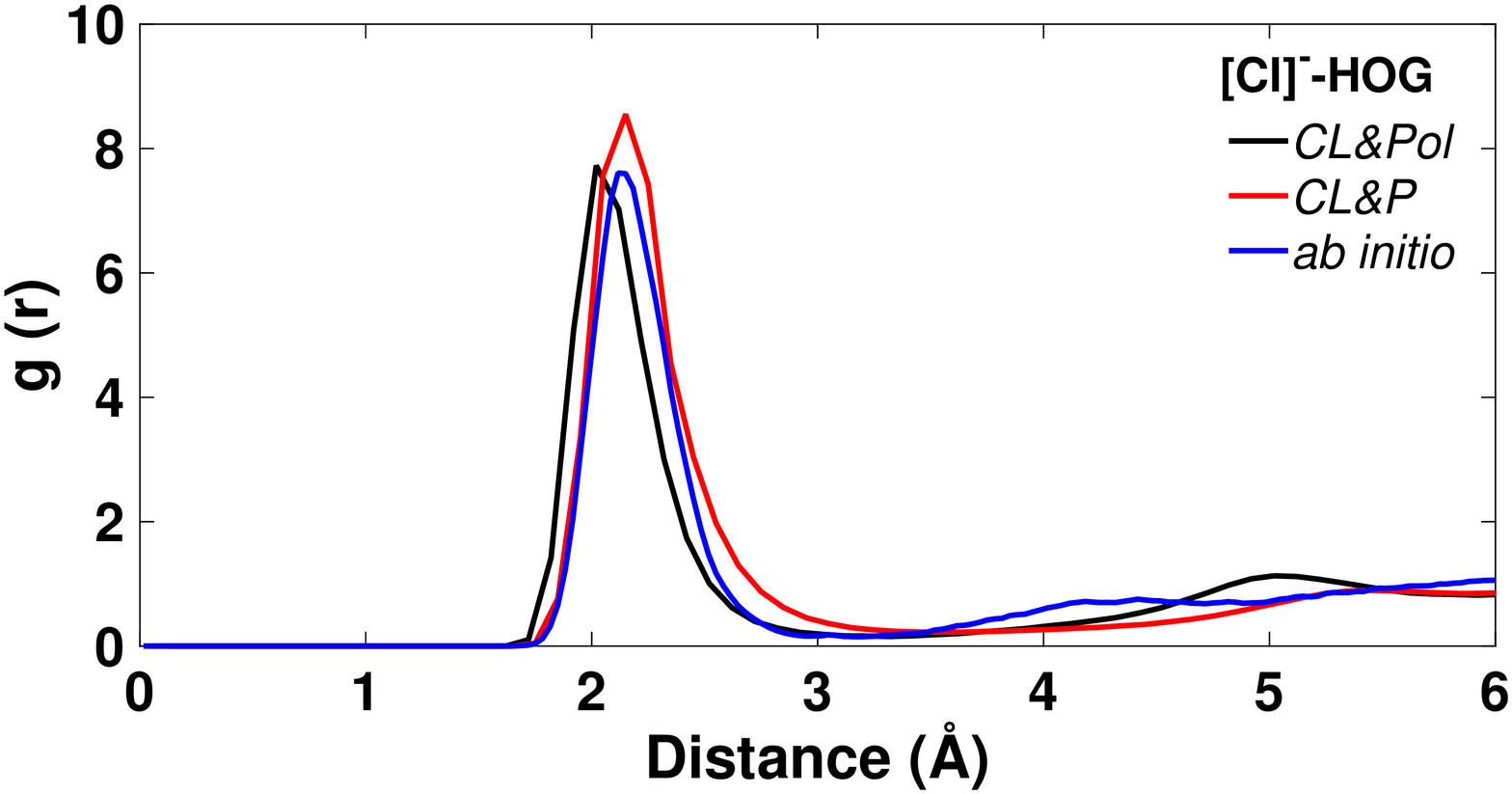}
\caption{Radial distribution functions between the \ce{[Cl]-} anions and the hydrogen atoms (see Figure{~}\ref{fig:structures} for nomenclature) of the hydroxyl groups at \SI{373}{\kelvin}. The \textit{ab initio} curves are from Ref.~\cite{Alizadeh2020}.}
\label{fig:grs}
\end{figure}

In the original publication of the CL{\&}Pol force field, the authors increased the repulsion ($\sigma$ parameter of the LJ potential) of the $\ce{[Cl]^-}-\ce{OH}$ and $\ce{[Cl]^-}-\ce{OHG}$ pairs to avoid the corresponding intense peaks in the radial distribution functions (RDF) due to the strong interactions that "froze" the system\cite{Goloviznina2021}. However, as we show in Figure{~}\ref{fig:phase} (left), using the original parameters, the ethaline system phase separates in an unphysical and unrealistic manner to  an ethylene glycol phase and a choline chloride phase. This phase separation occurred in approximately \SI{20}{\nano\second} at \SI{323.15}{\kelvin}. It is plausible that the original authors of the CL{\&}Pol force field did not observe that since their simulations were up to \SI{10}{\nano\second} and at \SI{298}{\kelvin}.  However, their RDF of the $\ce{[Cl]^-}-\ce{OHG}$ atom pair already provided a clue (see Figure~\num{3} in Goloviznina 
\textit{et al.}\cite{Goloviznina2021}):
The peak in the $g(r)$ was almost entirely absent indicating too weak interactions. 

The straightforward approach to correct the above issue would be 
to rescale the $\sigma$ parameters of the $\ce{[Cl]^-}-\ce{OH}$ and $\ce{[Cl]^-}-\ce{OHG}$ atom pairs to balance the interactions between the chlorides and the hydroxyls. Goloviznina \textit{et al.}\cite{Goloviznina2021} originally proposed $\sigma$=\SI{0.37}{\nano\meter} for both pairs, which are the values that drove the system to separate phases; they reported that with their original choice,
$\sigma$=\SI{0.337}{\nano\meter}, the interactions were too strong and the ethaline simulation froze\cite{Goloviznina2021}. The natural choice would be some value of $\sigma$ between \SI{0.337}{\nano\meter}-\SI{0.37}{\nano\meter}. In principle, there is no reason to adopt equal values for both $\ce{[Cl]^-}-\ce{OH}$ and $\ce{[Cl]^-}-\ce{OHG}$ atom pairs, since the LJ parameters were scaled differently to avoid double counting the effects of polarization, when transforming the CL{\&}P force field into the CL{\&}Pol force field. 

\begin{table}[t!]
\centering
\caption{Experimental\cite{Yadav2015, HarifiMood2017, DAgostino2011, Chen2019_T} (Exp) and calculated properties (at \SI{323.15}{\kelvin}) of ethaline using the polarizable CL{\&}Pol and the non-polarizable CL{\&}P force fields. Densities ($\rho$) are given in units of \SI{}{\gram\per\cubic\centi\metre}, viscosities ($\eta$) in \SI{}{cP^{-1}}, diffusion coefficients ($D^{+}$, $D^{0}$, $D^{-}$) in \SI{e-11}{\meter\squared\per\second} and surface tension ($\gamma$) in \SI{}{mN.m^{-1}}.}
\label{tbl:FF}
\setlength{\tabcolsep}{2.3pt}
\begin{tabular}{cccc} \toprule
      & Exp & CL{\&}Pol  & CL{\&}P  \\ \midrule
    $\rho$           & \tablenum{1.10} & \tablenum{1.114} ($\pm\num{0.002}$) & \tablenum{1.076} ($\pm\num{0.002}$) \\
    $\eta$           & \tablenum{18.8} & \tablenum{18.5} ($\pm\num{1.2}$) & \tablenum{19.7} ($\pm\num{1.2}$) \\
    $D^{+}$           & \tablenum{7.5} & \tablenum{4.6} ($\pm\num{0.5}$) & \tablenum{5.3} ($\pm\num{0.3}$) \\
    $D^{0}$           & \tablenum{13.2} & \tablenum{13.3} ($\pm\num{0.6}$) & \tablenum{10.4} ($\pm\num{0.4}$) \\
    $D^{-}$           & {-} & \tablenum{11.7} ($\pm\num{0.7}$) & \tablenum{7.9} ($\pm\num{0.5}$) \\
    $\gamma$           & \tablenum{47.6} & \tablenum{48.1} ($\pm\num{3.1}$) & \tablenum{48.0} ($\pm\num{2.9}$) \\ \bottomrule
\end{tabular}
\end{table}

To determine the appropriate $\sigma$ values for the $\ce{[Cl]^-}-\ce{OH}$ and $\ce{[Cl]^-}-\ce{OHG}$ pairs, our strategy was to reproduce the \textit{ab initio} reference RDFs at \SI{373.15}{\kelvin} available from Alizadeh \textit{et al.}\cite{Alizadeh2020}. We determined the values to be $\sigma$=\SI{0.345}{\nano\meter} and $\sigma$=\SI{0.356}{\nano\meter} for the $\ce{[Cl]^-}-\ce{OH}$ and $\ce{[Cl]^-}-\ce{OHG}$ pairs, respectively. The radial distribution functions are shown in Figure{~}\ref{fig:grs}. As the figure shows, the polarizable model is in excellent agreement with the \textit{ab initio} curves. The local structures of the $\ce{[Cl]^-}-\ce{OH}$ and $\ce{[Cl]^-}-\ce{OHG}$ correlations have maxima \num{9.9} at \SI{2.05}{Å} and \num{7.7} at \SI{2.0}{Å}. The corresponding \textit{ab initio} maxima are \num{9.8} at \SI{2.10}{Å} and \num{7.6} at \SI{2.12}{Å}. For comparison, the RDFs from the reduced charge CL{\&}P force field are also shown in Figure{~}\ref{fig:grs}. While positions of the the maxima are well reproduced, the peak intensities are suboptimal; the CL{\&}P force field predicts more intense peaks for the $\ce{[Cl]^-}-\ce{OHG}$ correlation than for the $\ce{[Cl]^-}-\ce{OH}$ correlation, while the CL{\&}Pol force field predicts the opposite, which corroborates with the \textit{ab initio} data\cite{Alizadeh2020}. 

Importantly, the new $\sigma$ parameters  correct the problem of the unphysical phase separation as illustrated in Figure{~}\ref{fig:phase} (right). The system remained stable (mixed) during the whole \SI{70}{\nano\second} simulation.  To check the robustness of this result, some basic properties (density, self-diffusion coefficients, viscosities and surface tension) of ethaline were calculated with both the polarizable and non-polarizable models and compared with available experimental data. We found good agreement with experimental data\cite{Yadav2015, HarifiMood2017, DAgostino2011, Chen2019_T}.
The results are summarized in Table~\ref{tbl:FF} and the details of the computations are given in Supporting Information. 

\subsection{Correcting Chlorides' Overpolarization}

\begin{figure}[t!]
\centering
\includegraphics[width=1.00\linewidth]{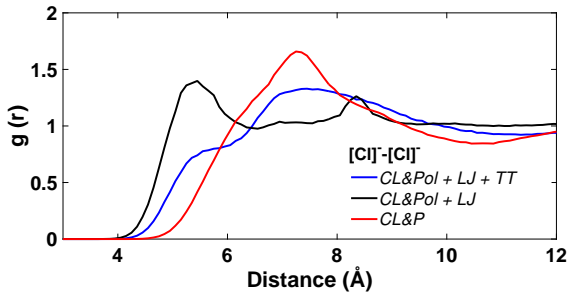}
\caption{Radial distribution functions between the \ce{[Cl]-}- \ce{[Cl]-} anions at \SI{373.15}{\kelvin}.
}
\label{fig:grs_cl}
\end{figure}

The \ce{[Cl]^-}-\ce{[Cl]^-} radial distribution functions (RDF) are shown in Figure{~}\ref{fig:grs_cl}. For comparison, the curve for the non-polarizable CL{\&}P force field is also included. The black curve shows the result with direct application of the fine-tuned LJ parameters of the CL{\&}Pol force field from the previous section. Two peaks, a principal peak at \num{5.5}\,Å and a secondary peak at \num{8.3}\,Å are present. In contrast, the non-polarizable force field presents only one peak of relatively large width with the maximum at \num{7.2}\,Å.  The blue curve shows the result for the CL{\&}Pol force field with both LJ parameters and overpolarization (using the TT damping function) corrected and we will discuss it next.

We use the  \textit{ab initio} RDF data of Alizadeh \textit{et al.}\cite{Alizadeh2020} as a reference (Figure~2 in their article). Their data is very similar to the
non-polarizable CL{\&}P data, the red curve in Figure{~}\ref{fig:grs_cl}. This comparison indicates that the CL{\&}P force field correctly reproduces the RDF while the polarizable CL{\&}Pol force field does not do so even after the LJ parameter correction (black curve in Figure{~}\ref{fig:grs_cl}). At first sight that is somewhat surprising.

Szabadi \textit{et al.}\cite{Szabadi2021} compared \textit{ab initio} and polarizable MD simulations of aqueous chloride-based ionic liquids and found that all of the investigated polarizable force fields (including CL{\&}Pol) overestimate the induced dipoles of the chlorides due to their high polarizability, $\alpha_{Cl}$=\num{4.4}\,{\AA}$^{3}$. 
The strong induced dipoles counteract the Coulomb repulsion between the chlorides. Consequently, Szabadi \textit{et al.} observed\cite{Szabadi2021} an alignment of chlorides with water molecules and formation of aggregates that would be unfavorable in the presence of properly balanced interactions.  They attempted rectify this overpolarization by reducing chlorides' polarizability\cite{Szabadi2021}. Although the results improved, there were still inconsistencies and the use of the TT damping function was suggested\cite{Szabadi2021}. 

To correct, or at least to attenuate, the above issue, we applied the TT damping function\cite{Tang1984} as suggested by Szabadi \textit{et al.}\cite{Szabadi2021}. This approach is simple and takes advantage of the fact that the TT potential is already included in the CL{\&}Pol force field definition\cite{Goloviznina2021} and implemented in LAMMPS. The RDF after applying this correction is shown as the blue curve in Figure{~}\ref{fig:grs_cl}. 
The agreement with the non-polarizable curve (red), and hence the \textit{ab initio} reference data\cite{Alizadeh2020}, is much better: With the correction, the \ce{[Cl]^-}-\ce{[Cl]^-} RDF has only one peak at the expected position, although the peak is slightly wider and less sharp.

\begin{figure}[b!]
\centering
\includegraphics[width=1.00\linewidth]{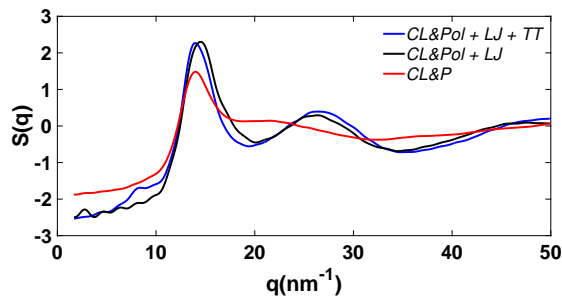}
\caption{Total X-ray structure factors, $S(q)$, for bulk ethaline at \SI{323.15}{\kelvin}
computed using Equation~\ref{eqn:TT}.
}
\label{fig:SK_Total}
\end{figure}

\begin{figure*}[t!]
\centering
\begin{minipage}[!]{0.49\linewidth}
\includegraphics[width=1.00\linewidth]{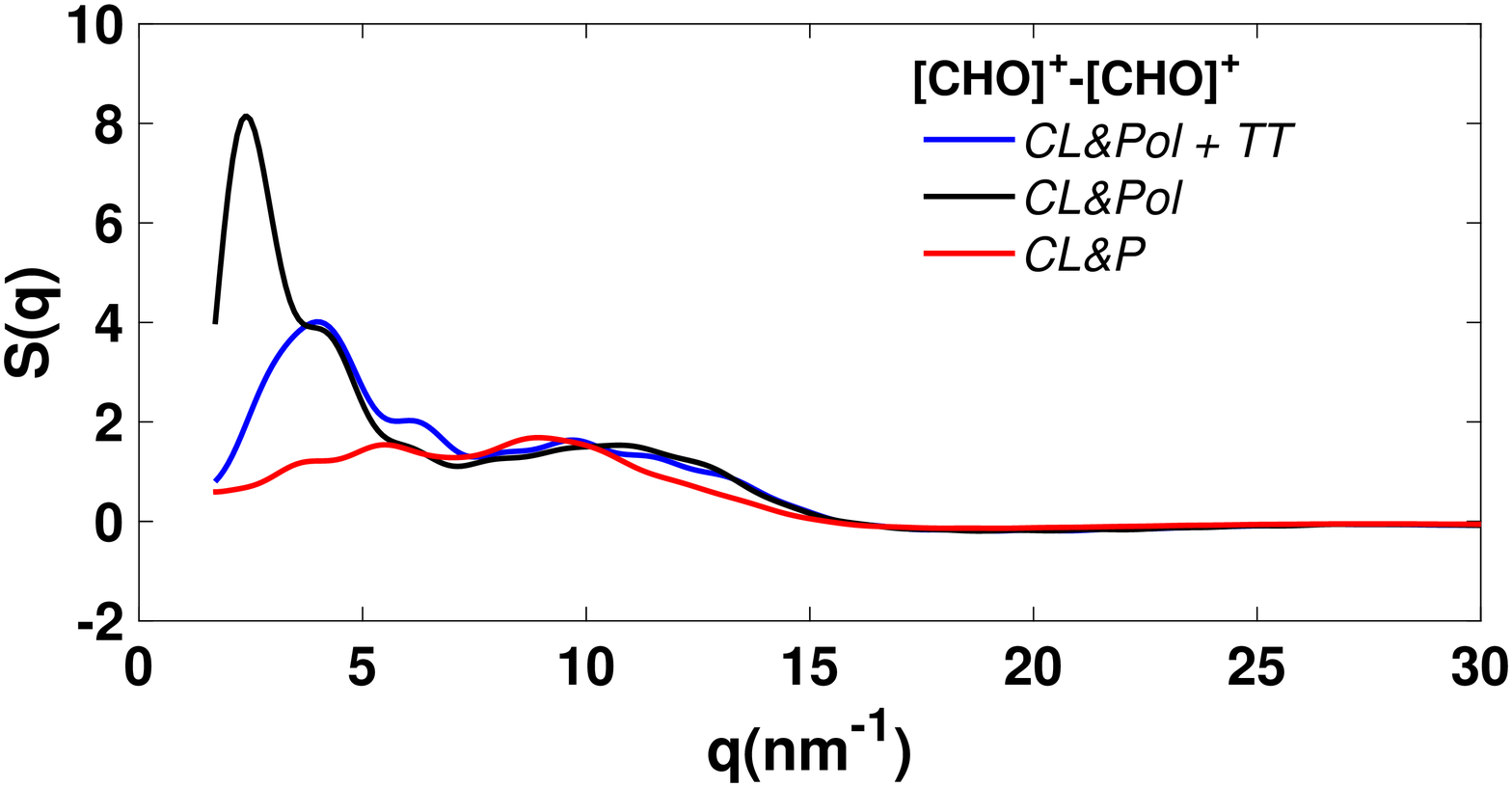}
\end{minipage}
\begin{minipage}[!]{0.49\linewidth}
\includegraphics[width=1.00\linewidth]{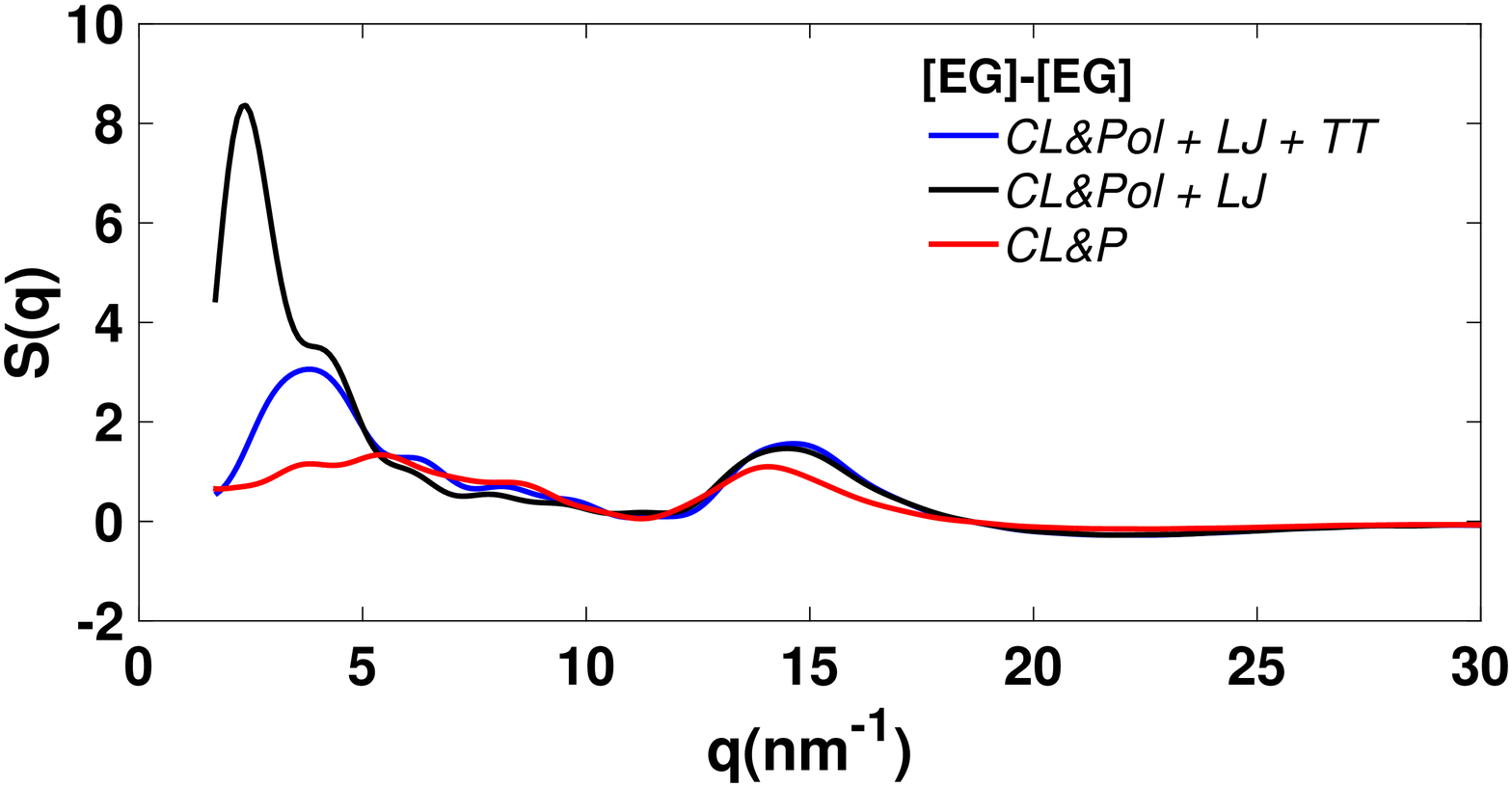}
\end{minipage}
\begin{minipage}[!]{0.49\linewidth}
\includegraphics[width=1.00\linewidth]{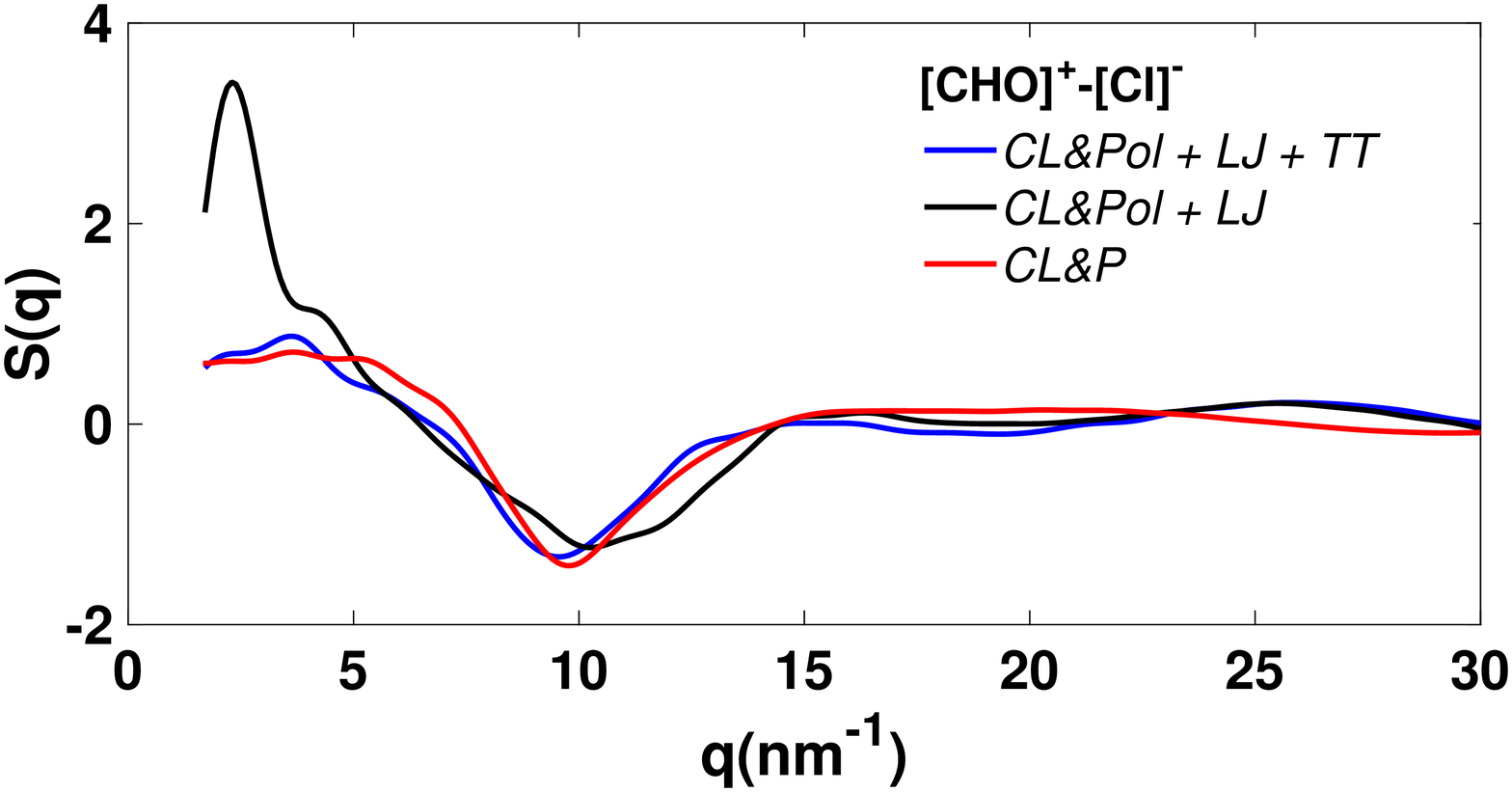}
\end{minipage}
\begin{minipage}[!]{0.49\linewidth}
\includegraphics[width=1.00\linewidth]{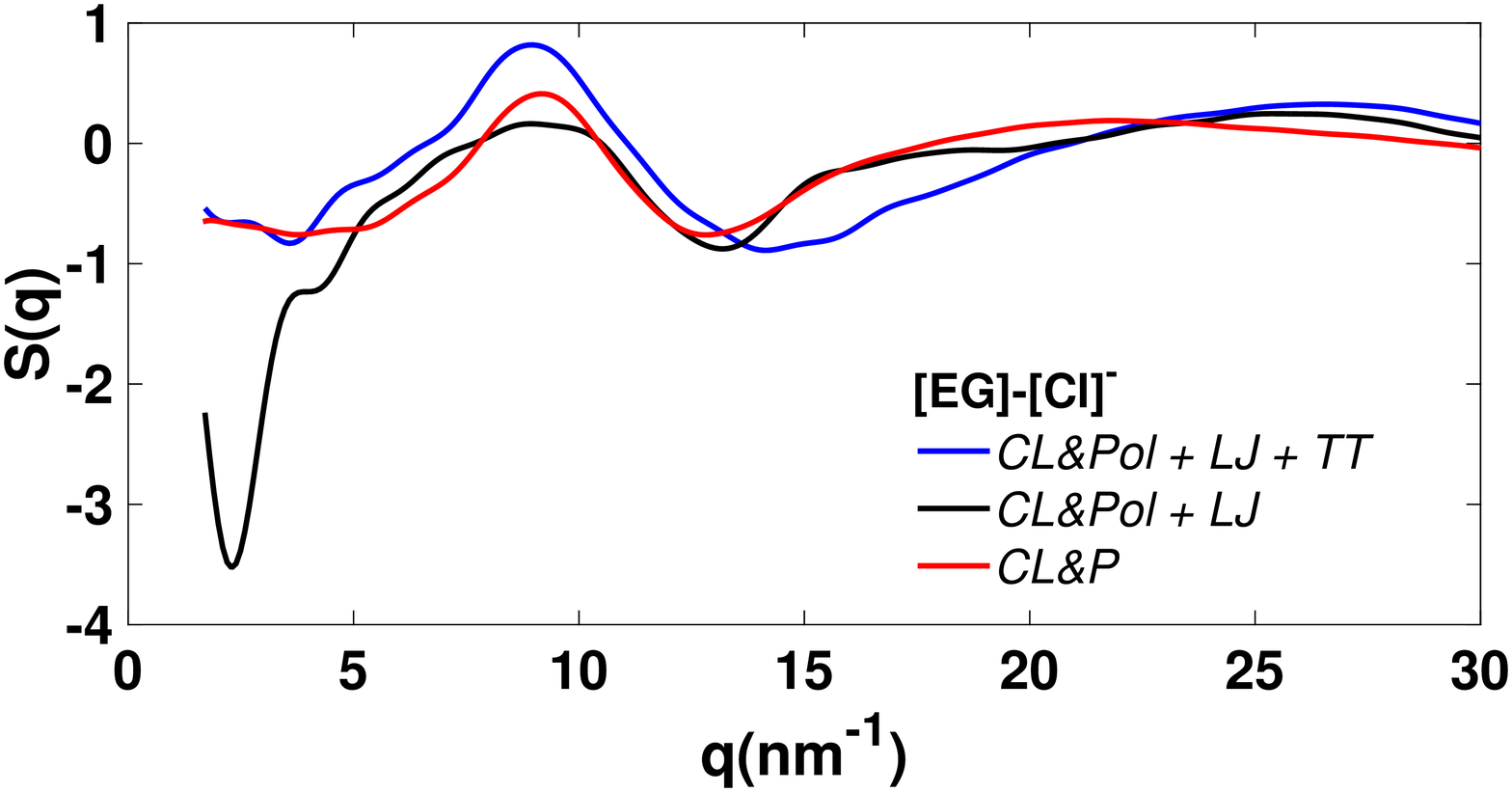}
\end{minipage}
\begin{minipage}[!]{0.49\linewidth}
\includegraphics[width=1.00\linewidth]{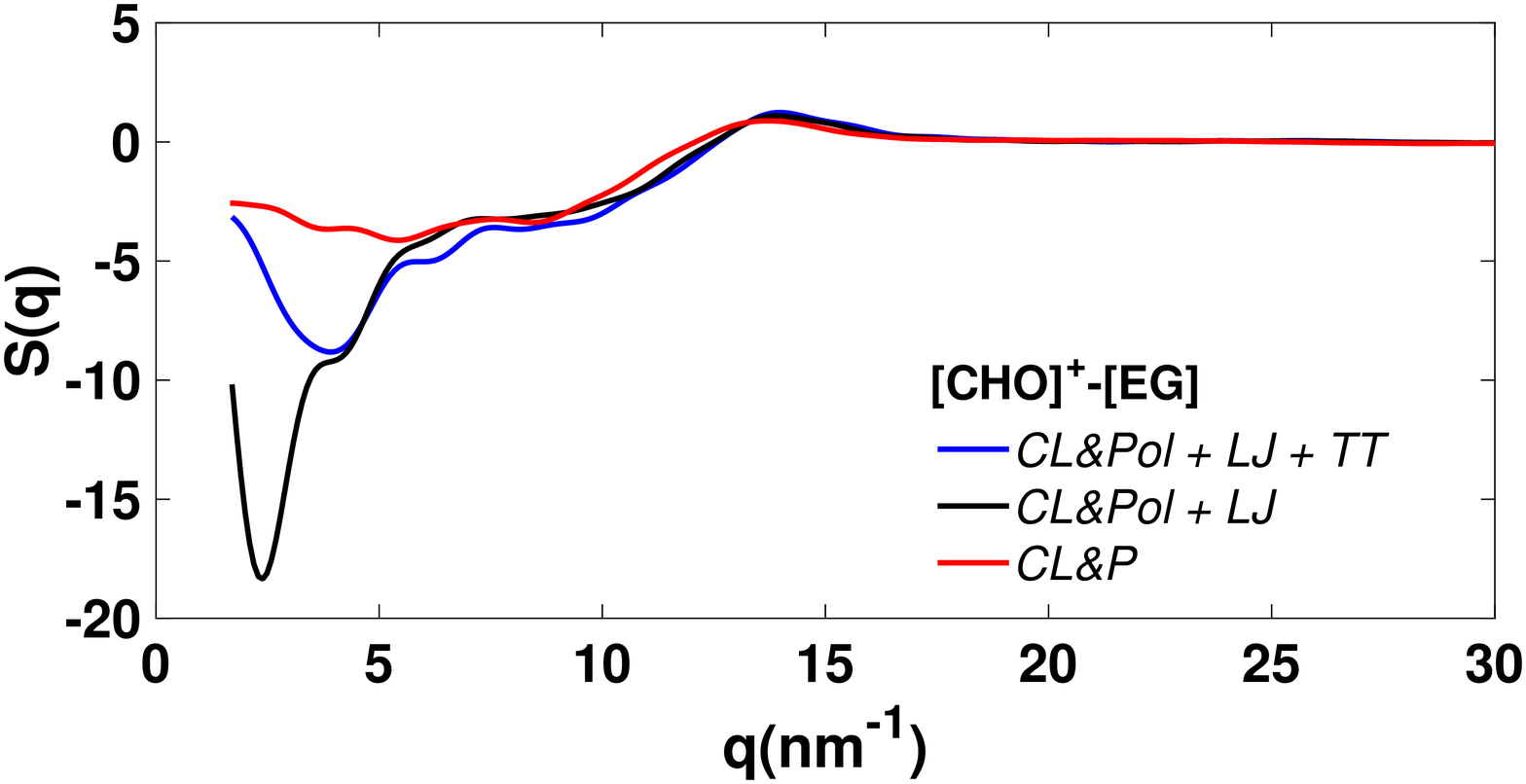}
\end{minipage}
\begin{minipage}[!]{0.49\linewidth}
\includegraphics[width=1.00\linewidth]{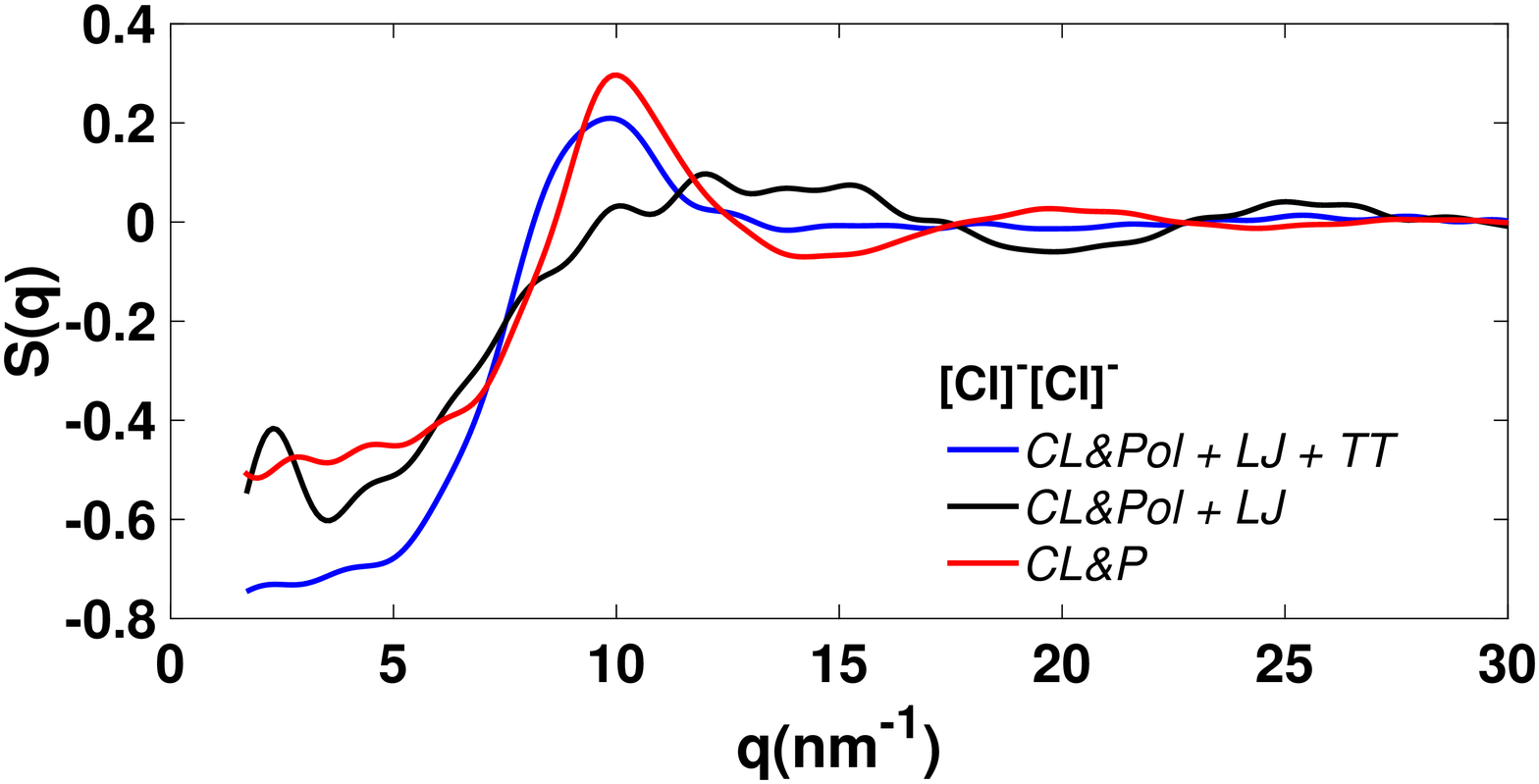}
\end{minipage}
\caption{Partial X-ray structure factors, $S(q)$, for bulk ethaline correlation components at \SI{323.15}{\kelvin}
computed using Equation~\ref{eqn:SK_Partial}.
}
\label{fig:SKs}
\end{figure*}

A deeper structural analysis was done by computing the total and the partial components of the X-ray structure factor\cite{Sharma2021}, $S(q)$, with the Travis\cite{Brehm2011, Brehm2020} software using
\begin{center}
\begingroup
  \small
  \thinmuskip=\muexpr\thinmuskip*1/9\relax
  \medmuskip=\muexpr\medmuskip*1/9\relax 
\begin{equation}\label{eqn:TT}
\begin{split}
S(q) = \frac{\rho_{0}\sum_{i=1}^{n}\sum_{j=1}^{n}x_{i}x_{j}f_{i}(q)f_{j}(q)} 
{[\sum_{i=1}^{n}x_{i}f_{i}(q)][\sum_{j=1}^{n}x_{j}f_{j}(q)]
} \\ \times
\int_{0}^{L/2}\num{4}\pi r^{2}[g_{ij}(r) - \num{1}] \frac{\sin{qr}}{qr}W(r)\,dr,
\end{split}
\end{equation}
\endgroup\\
\end{center}
where $\rho_{0}$ is the total number density, $x_{i}$ and $x_{j}$ are the molar fractions of atoms $i$ and $j$, $f_{i}(q)$ and $f_{j}(q)$ are tabulated X-ray atomic form factors, $L$ is the simulation box length, $g_{ij}(r)$ is the radial distribution function between atomic species including both intra- and intermolecular terms, and $W(r) = \sin(2\pi r/L)/(2\pi r/L)$ is a Lorch function, sometimes used to reduce the effects of finite truncation error of $g_{ij}(r)$ at large values of $r$.  The partial components of $S(q)$ were computed using\cite{Kaur2019, Kaur2020}
\begin{center}
\begingroup
  \small   
  \thinmuskip=\muexpr\thinmuskip*1/9\relax
  \medmuskip=\muexpr\medmuskip*1/9\relax 
\begin{equation}\label{eqn:SK_Partial}
\begin{split}
S(q) = S^{[CHO]^+-[CHO]^+}(q) + S^{[Cl]^--[Cl]^-}(q) \\ 
+ S^{[EG]-[EG]}(q) + \num{2}S^{[CHO]^+-[Cl]^-}(q) \\ 
+ \num{2}S^{[CHO]^+-[EG]}(q) 
+ \num{2}S^{[Cl]^--[EG]}(q). \\ \end{split}
\end{equation}
\endgroup\\
\end{center}

The total X-ray structure factors for ethaline are shown in Figure{~}\ref{fig:SK_Total}.
There is a principal peak around \SI{14}{\per\nano\meter}, which corresponds to a distance of approximately $2\pi/14 \! \approx \! \SI{0.44}{\nano\meter}$. Besides that, the secondary peak at \SI{26}{\per\nano\meter} and the valley at \SI{35}{\per\nano\meter} are more prominent with the polarizable model, both with and without the overpolarization correction. In turn, the non-polarizable CL{\&}P model shows total X-ray structure factors slightly shifted toward the lower $q$-vector region (higher real space values). 

In general, the main contributions to the principal peak are from the \ce{[EG]}-\ce{[EG]} and \ce{[CHO]^+}-\ce{[EG]} correlations, as the partial components of $S(q)$ in Figure{~}\ref{fig:SKs} show. Another interesting observation is the presence of peaks and antipeaks in the different self- and cross-correlations around \SI{10}{\per\nano\meter} suggesting a pseudo-charge-ordering similar to what has been found for ILs\cite{Sharma2021, Hettige2012} and also previously reported for DESs\cite{Kaur2019, Kaur2020}. Those peaks are due to \ce{[CHO]^+}-\ce{[CHO]^+}, \ce{[EG]}-\ce{[Cl]^-}, and \ce{[Cl]^-}-\ce{[Cl]^-} correlations, while the antipeaks are mainly due to \ce{[CHO]^+}-\ce{[Cl]^-} correlation.

\begin{figure*}[t!]
\centering
\includegraphics[width=1.00\linewidth]{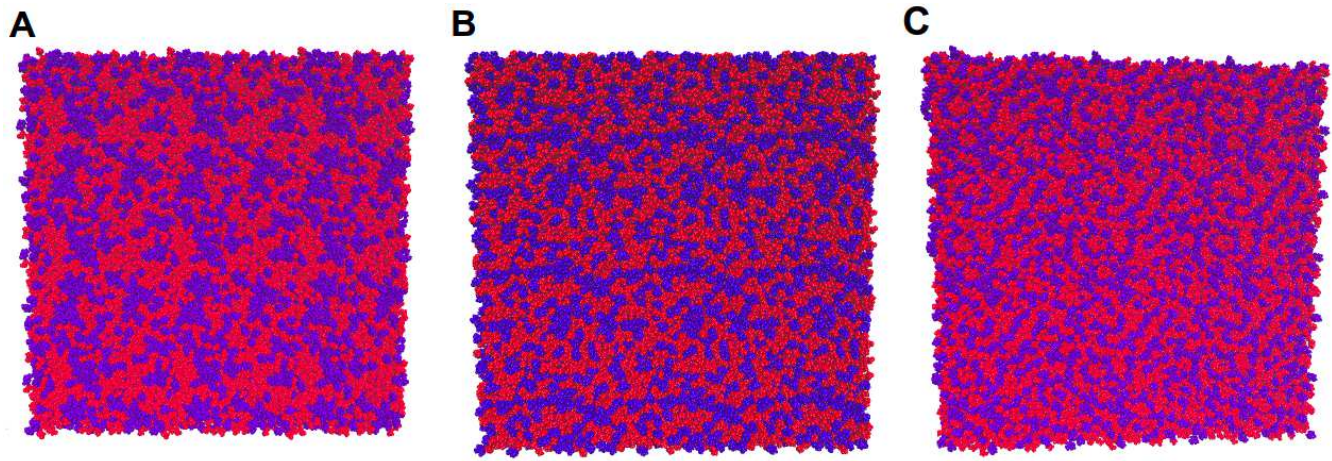}
\caption{Representative snapshot of ethaline simulated with (A) CL{\&}Pol, (B) CL{\&}Pol + LJ + TT, and (C) CL{\&}P + LJ. The images illustrate the degree of long-range ordering in the system. Choline ions and ethylene glycol molecules are depicted in blue and red, respectively.
}
\label{fig:nanohetero}
\end{figure*}

At first sight the overpolarization correction does not seem to have a considerable impact on the liquid structure in Figure{~}\ref{fig:SK_Total}. However, analysis of the partial components of $S(q)$ shows remarkable differences, as seen in Figure{~}\ref{fig:SKs}. The first one concerns the \ce{[Cl]^-}-\ce{[Cl]^-} correlation, as may be expected based on the \ce{[Cl]^-}-\ce{[Cl]^-} RDFs. Without the overpolarization correction, there is more than one peak in the range between \num{10}-\SI{15}{\per\nano\meter}. In addition, the peaks are considerably smaller than the single peak around \SI{10}{\per\nano\meter} for the CL{\&}P force field and the corrected CL{\&}Pol force field.

When overpolarization is present, the notable difference is the presence of prepeaks and preantipeaks at \SI{2.4}{\per\nano\meter} in all self- and cross-correlations. This corresponds to a distance of  $2\pi/2.4 \!\! \approx \! \! \SI{2.93}{\nano\meter}$, indicating long-range structural ordering or nano-heterogeneities. In contrast, when overpolarization is corrected, there are less intense prepeaks and preantipeaks and they appear only in the \ce{[CHO]^+}-\ce{[CHO]^+}, \ce{[EG]}-\ce{[EG]}, and \ce{[CHO]^+}-\ce{[EG]} correlations. In addition, the position of the prepeaks and preantipeaks change to \SI{4.3}{\per\nano\meter}, corresponding to a distance of $2\pi/4.3 \! \approx \! \SI{1.46}{\nano\meter}$. 
This is approximately half of the distance when the overpolarization is present. The non-polarizable model does not show any prepeaks or preantipeaks at low $q$ values, that is, no long-range ordering. 

Aiming to understand the differences of long-range ordering between the models representative snapshots of the systems (with some periodic copies included) are shown in Figure{~}\ref{fig:nanohetero}. In the case of the CL{\&}Pol force field without overpolarizarion correction (Fig.~\ref{fig:nanohetero}A), the presence of choline-rich (blue) and  ethylene glycol-rich (red) nano-heterogeneities is evident. In the case of the non-polarizable model (Fig.~\ref{fig:nanohetero}C), the choline ions and ethylene glycol molecules are uniformly spread throughout the simulation box, without any noticeable domains. In turn, when overpolarization is corrected in the CL{\&}Pol force field (Fig.~\ref{fig:nanohetero}B), the situation is intermediate to the other two cases. In summary, Figure{~}\ref{fig:nanohetero} gives a visual analysis of the aforementioned discussion of prepeaks and preantipeaks at very low $q$ region.

\begin{table*}[t!]
\centering
\caption{Average surface area (\SI{}{Å^2}) shared between components of ethaline at \SI{323.15}{\kelvin} obtained through Voronoi tesselation. The entries show the average area which a \textit{single} ion/molecule shares with the other component of the contact pair. The standard deviation is equal to \SI{0.1}{Å^2}.}
\label{tbl:Voro}
\begin{tabular}{cccc} \toprule
    Contact Pair  & CL{\&}Pol + LJ + TT & CL{\&}Pol + LJ  & CL{\&}P  \\ \midrule
    $\ce{[Cl]^-}-\ce{[Cl]^-}$           & \tablenum{0} & \tablenum{0} & \tablenum{0} \\
    $\ce{[Cl]^-}-\ce{[CHO]^+}$           & \tablenum{19.9} & \tablenum{20.7} & \tablenum{19.5} \\
    $\ce{[Cl]^-}-\ce{[EG]}$           & \tablenum{16.9} & \tablenum{14.9} & \tablenum{17.2} \\
    $\ce{[CHO]^+}-\ce{[CHO]^+}$           & \tablenum{409.1} & \tablenum{420.5} & \tablenum{404.6} \\
    $\ce{[CHO]^+}-\ce{[EG]}$           & \tablenum{108.6} & \tablenum{95.1} & \tablenum{113.7} \\
    $\ce{[EG]}-\ce{[EG]}$           & \tablenum{211.4} & \tablenum{219.5} & \tablenum{208.5} \\ \bottomrule
\end{tabular}
\end{table*}

A few MD simulations of ethaline have been performed before\cite{Alizadeh2020, Ferreira2016, Zhang2020, Kaur2019}. However, the X-ray scattering structure factors were computed only by Kaur \textit{et al.}\cite{Kaur2019}, who also found peaks and antipeaks in low $q$ region, around \SI{5}{\per\nano\meter} corresponding to a  distance of $2\pi/5 \approx \SI{1.25}{\nano\meter}$. This indicates some degree of nano-heterogeneity similar to what we have obtained after the overpolarization correction with the TT damping function. Nano-heterogeneity was also found by Zhang \textit{et al.}\cite{Zhang2020}, who computed protiated and deutered neutron scattering structure factors using the Genarlized Amber Force Field\cite{Wang2004, Sprenger2015} (GAFF) with scaled charges  and compared with experimental data. Although the match between simulation and experiment was not perfect, it was still good and showed specific structural correlations at all length scales, including the low $q$ region. Another interesting work was performed by Alizadeh \textit{et al.}\cite{Alizadeh2019}, who simulated choline chloride and some of its derivatives with elongated alkyl chains in the presence of ethylene glycol molecules. They also found noticeable heterogeneity across the systems. In addition, the larger the alkyl chain, the stronger the heterogeneity. In general, as reviewed by Kaur \textit{et al.}\cite{Kaur2020}, DESs should exhibit some heterogeneity at nanoscales, similar of the ILs. 

Considering the results in Figures~\ref{fig:SKs} and \ref{fig:nanohetero} and the above discussion, MD simulations of ethaline should show some, albeit limited, degree of nano-heterogeneity. However, nano-heterogeneity is overestimated in the CL{\&}Pol force field, as seen in Figure{~}\ref{fig:nanohetero}A. As will be discussed below, \ce{[CHO]^+}-rich complexes are preferentially formed due to the over-strong induced chloride dipoles. When overpolarization is corrected, \ce{[EG]}-rich complexes are preferentially formed. Both are, however, mediated by chlorides.

To quantify the above, the Voronoi tessellation\cite{Voronoi1908} technique was applied using the Voro++ package\cite{Rycroft2009} implemented in LAMMPS; Voronoi cell defines the area (volume) that is closer to a given particle than any other particle. Since the Voronoi cell of each molecule shares facets with other molecules, an average contact area between components can be computed. For this calculation, we removed all the Drude particles to allow for a direct comparison with the non-polarizable model as otherwise there would be artificial
facets due to the Drude particles and hence systematically higher areas. The results are presented in Table~\ref{tbl:Voro}.

\begin{figure}[b!]
\centering
\includegraphics[width=1.00\linewidth]{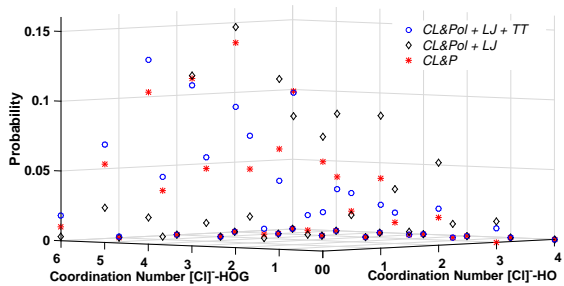}
\caption{Probability of simultaneous coordination of different amount of hydroxyls from choline and from ethylene glycol around a single chloride anion.
}
\label{fig:prob_clusters}
\end{figure}

\begin{table*}[t!]
\centering
\caption{Probabilities (\textbf{P}) of different coordination numbers (\textbf{CN}) of hydroxyls simultaneously coordinated with a single \ce{[Cl]^-} anion.}
\label{tbl:ProbCN}
\begin{tabular}{ccccc} \toprule
     \textbf{CN}  & \textbf{CN} & \textbf{P(\%)} & \textbf{P(\%)}  & \textbf{P(\%)}  \\ \midrule
     $\ce{[Cl]^-}-\ce{HO}$  & $\ce{[Cl]^-}-\ce{HOG}$ & CL{\&}Pol + LJ + TT & CL{\&}Pol + LJ  & CL{\&}P  \\ \midrule
    \tablenum{0}           & \tablenum{0} & \tablenum{2.8} & \tablenum{8.2} & \tablenum{6.4} \\
    \tablenum{0}           & \tablenum{1} & \tablenum{4.9} & \tablenum{12.2} & \tablenum{7.2} \\
    \tablenum{0}           & \tablenum{2} & \tablenum{10.1} & \tablenum{15.8} & \tablenum{14.7} \\
    \tablenum{0}           & \tablenum{3} & \tablenum{11.5} & \tablenum{12.2} & \tablenum{12.0} \\
    \tablenum{0}           & \tablenum{4} & \tablenum{13.2} & \tablenum{1.9} & \tablenum{10.9} \\
    \tablenum{0}           & \tablenum{5} & \tablenum{7.0} & \tablenum{2.5} & \tablenum{5.6} \\
    \tablenum{0}           & \tablenum{6} & \tablenum{1.8} & \tablenum{0.3} & \tablenum{1.0} \\
    \tablenum{1}           & \tablenum{0} & \tablenum{3.1} & \tablenum{9.5} & \tablenum{5.0} \\
    \tablenum{1}           & \tablenum{1} & \tablenum{4.1} & \tablenum{9.5} & \tablenum{5.0} \\
    \tablenum{1}           & \tablenum{2} & \tablenum{10.9} & \tablenum{9.2} & \tablenum{11.0} \\
    \tablenum{1}           & \tablenum{3} & \tablenum{7.7} & \tablenum{1.9} & \tablenum{5.3} \\
    \tablenum{1}           & \tablenum{4} & \tablenum{6.0} & \tablenum{1.3} & \tablenum{5.2} \\
    \tablenum{1}           & \tablenum{5} & \tablenum{4.5} & \tablenum{0.2} & \tablenum{3.5} \\
    \tablenum{1}           & \tablenum{6} & \tablenum{0.1} & \tablenum{0.0} & \tablenum{0.0} \\
    \tablenum{2}           & \tablenum{0} & \tablenum{2.6} & \tablenum{5.9} & \tablenum{2.0} \\
    \tablenum{2}           & \tablenum{1} & \tablenum{2.2} & \tablenum{3.9} & \tablenum{1.5} \\
    \tablenum{2}           & \tablenum{2} & \tablenum{3.5} & \tablenum{1.9} & \tablenum{2.2} \\
    \tablenum{2}           & \tablenum{3} & \tablenum{1.8} & \tablenum{0.4} & \tablenum{0.7} \\
    \tablenum{2}           & \tablenum{4} & \tablenum{0.7} & \tablenum{0.0} & \tablenum{0.3} \\
    \tablenum{2}           & \tablenum{5} & \tablenum{0.0} & \tablenum{0.0} & \tablenum{0.0} \\
    \tablenum{2}           & \tablenum{6} & \tablenum{0.0} & \tablenum{0.0} & \tablenum{0.0} \\
    \tablenum{3}           & \tablenum{0} & \tablenum{1.0} & \tablenum{1.5} & \tablenum{0.0} \\
    \tablenum{3}           & \tablenum{1} & \tablenum{0.2} & \tablenum{1.2} & \tablenum{0.2} \\
    \tablenum{3}           & \tablenum{2} & \tablenum{0.3} & \tablenum{0.5} & \tablenum{0.3} \\
    \tablenum{3}           & \tablenum{3} \tablenum{4} \tablenum{5} \tablenum{6} & \tablenum{0.0} & \tablenum{0.0} & \tablenum{0.0} \\
    \tablenum{4}           &\tablenum{0} \tablenum{1} \tablenum{2} \tablenum{3} \tablenum{4} \tablenum{5} \tablenum{6} & \tablenum{0.0} & \tablenum{0.0} & \tablenum{0.0} \\ \bottomrule
\end{tabular}
\end{table*}

The average surface areas of the
\ce{[CHO]^+}-\ce{[CHO]^+} and \ce{[EG]}-\ce{[EG]} contact pairs are \SI{420.5}{Å^2} and \SI{219.5}{Å^2}, respectively, for the CL{\&}Pol force field with overpolarization. These values are larger than the corresponding values of \SI{409.1}{Å^2} and \SI{211.4}{Å^2} from the CL{\&}Pol force field with overpolarization corrected. The values are larger due to the overestimated segregation of the long-range ordering, as seen in Figure{~}\ref{fig:nanohetero}. Besides that, for the CL{\&}Pol force field with overpolarization, the chlorides share more area with the choline cations (\SI{20.7}{Å^2}) and less area with the ethylene glycol molecules (\SI{14.9}{Å^2}) when compared to the CL{\&}P (\SI{19.5}{Å^2} ; \SI{17.2}{Å^2}) and the CL{\&}Pol with the overpolarization corrected (\SI{19.9}{Å^2} ; \SI{16.9}{Å^2}). As suggested by these values, when overpolarization is present chloride-mediated \ce{[CHO]^+} complexes are favored due to larger \ce{[Cl]^-}-\ce{[CHO]^+} and \ce{[CHO]^+}-\ce{[CHO]^+} areas, while when overpolarization is corrected chloride-mediated \ce{[EG]} complexes are favored.

To confirm the above, we calculated the probabilities of different amounts of hydroxyls from \ce{[CHO]^+} and \ce{[EG]} species simultaneously coordinated to a single \ce{[Cl]^-} anion. Each \ce{[CHO]^+} cation contributes with up to \num{1} hydroxyl group and each \ce{[EG]} molecule contributes with up to \num{2} hydroxyls. The criteria to assign a coordination is the same that determines a standard hydrogen bond with default values adopted from the Visual Molecular Dynamics (VMD) software\cite{Humphrey1996}. The results are graphically presented in Figure{~}\ref{fig:prob_clusters} and each probability given in Table~\ref{tbl:ProbCN}. 

Most of the points in Figure{~}\ref{fig:prob_clusters} have very low probabilities or even zero. These probabilities correspond to the case where the \ce{[Cl]^-} ions would be simultaneously coordinated with several hydroxyls. In this case, the hydroxyls can not achieve, at the same time, the proper conditions of distance and directionality to coordinate around the same chloride. In turn, many of the points with higher probabilities for the CL{\&}Pol force field with and without the overpolarization correction are located in distinct regions of Figure{~}\ref{fig:prob_clusters}:
Middle top to right (black points) and middle top to left (blue points). These point location regions reflect the following: As the number of \ce{HOG} groups coordinated to a \ce{[Cl]^-} anion increases, the probabilities get higher for the CL{\&}Pol force field with overpolarization corrected than for the CL{\&}Pol without overpolarization correction. In contrast, when the number of \ce{HOG} groups decreases or when the number of \ce{HO} groups increases, the probabilities get lower for the CL{\&}Pol force field with overpolarization corrected and higher without correction. For instance, consulting Table~\ref{tbl:ProbCN}, when $\ce{[Cl]^-}-\ce{HO}$ $= 1$ and $\ce{[Cl]^-}-\ce{HOG}$ $= 3$ the probability is \num{7.7}$\%$ and \num{1.3}$\%$ with and without overpolarization correction, respectively. On the other hand, when $\ce{[Cl]^-}-\ce{HO}$ $= 1$ and $\ce{[Cl]^-}-\ce{HOG}$ $= 1$ the probabilities are \num{4.1}$\%$ and \num{9.5}$\%$, respectively. In the end, considering that there are many possible combinations of single-chloride mediated complexes, the impact on the summed probabilities is not negligible, that is, \ce{[EG]}-rich complexes are preferentially found when the overpolarization is corrected and \ce{[CHO]^+}-rich complexes are favored when the overpolarization is not corrected, corroborating with the discussions from the average surface area in Table~\ref{tbl:Voro} and with the overestimated nano-heterogeneity found from the low $q$ region of Figure{~}\ref{fig:SKs}.

\section{Conclusion}

In this work, we focused on correcting two problems that are present in the polarizable CL{\&}Pol force field for the deep eutectic solvent ethaline\cite{Goloviznina2019,Goloviznina2021}. Our simulations showed 1) unphysical phase separation in long simulations and 2) appearance of artificial nano-scale heterogeneities. Two different corrections were needed, the first involving the Lennard-Jones parameters and the second the inclusion of the Tang-Toennis damping function\cite{Tang1984} to correct for overpolarization.

The first problem has its origin in the so-called "naked" hydrogens, that is, hydrogen atoms that do not have LJ parameters. This leads to unrealistically strong interactions after addition of the Drude particles.
To correct for this, we balanced the interactions of the \ce{[Cl]^-} anions with the hydroxyls from the \ce{[CHO]^-} cation and \ce{[EG]^-} molecules. The values we propose here are $\sigma$=\SI{0.345}{\nano\meter} and $\sigma$=\SI{0.356}{\nano\meter} for the $\ce{[Cl]^-}-\ce{OH}$ and $\ce{[Cl]^-}-\ce{OHG}$ pairs, respectively. With these, the \textit{ab initio} reference RDFs\cite{Alizadeh2020} were reproduced and no artificial phase separation occurred. We would like to add a word of caution, however: When simulating other DESs, it is not clear that the $\sigma$ parameters are directly transferable.

The second issue, overpolarization of chlorides, has its physical origin in the high polarizability of chloride, $\alpha_{Cl}$=\num{4.4}\,{\AA}$^{3}$. 
When not properly corrected, overpolarization leads to the appearance of unphysical
nano-scale heterogeneities manifested as prepeaks and preantipeaks at \SI{2.4}{\per\nano\meter} in all self- and cross-correlations of the partial structure factors. We corrected the overpolarization  by applying the Tang-Toennis damping function\cite{Tang1984} in the interactions of the induced chloride dipoles. This was originally suggested by Szabadi \textit{et   al.}\cite{Szabadi2021} but not applied. The correction took care of artificial structural heterogeneity and the \ce{[Cl]^-}-\ce{[Cl]^-} RDF became similar to the \textit{ab initio} reference curve. 

\section{Data and Software Availability}

All the force field parameters for ethaline are available in github from the developers of the CL{\&}Pol force field. (\url{https://github.com/kateryna-goloviznina/desff/tree/master/example_pol-des}). The updated values for the $\sigma$ parameter of the $\ce{[Cl]^-}-\ce{OHG}$ and $\ce{[Cl]^-}-\ce{OH}$ interactions are $\sigma$=\SI{0.345}{\nano\meter} and $\sigma$=\SI{0.356}{\nano\meter}, respectively. 
All software used in this work (LAMMPS, VMD, Travis, and Packmol) is freely available on the internet. Structures of the systems or raw MD data are available upon request.

\begin{acknowledgement}
R.M.d.S. and M.C.C.R. thank FAPESP (The S\~{a}o Paulo Research Foundation) grants Process 2020/06766-9 and 2016/21070-5. M.K. thanks the Natural Sciences and Engineering Research Council of Canada (NSERC)
and the Canada Research Chairs Program. Computational resources were provided by USP High Performance Computing (USP-HPC), SDumont (https://sdumont.lncc.br) from the "Laboratório Nacional de Computação Científica (LNCC/MCTI, Brazil)", and Compute Canada (www.computecanada.ca).
This research was funded by the Ministry of Education and Science of the Russian Federation (contract RF----225121X0043).
R.M.d.S. also thanks Vahideh Alizadeh and Barbara Kirchner for sharing the \textit{ab initio} data from Figure{~}\ref{fig:grs}.
\end{acknowledgement}

\begin{suppinfo}
Details of how density, viscosity, self-diffusion coefficients and surface tension were computed.
\end{suppinfo}


\providecommand{\latin}[1]{#1}
\makeatletter
\providecommand{\doi}
  {\begingroup\let\do\@makeother\dospecials
  \catcode`\{=1 \catcode`\}=2 \doi@aux}
\providecommand{\doi@aux}[1]{\endgroup\texttt{#1}}
\makeatother
\providecommand*\mcitethebibliography{\thebibliography}
\csname @ifundefined\endcsname{endmcitethebibliography}
  {\let\endmcitethebibliography\endthebibliography}{}

\end{document}